\documentclass[conference]{IEEEtran}
\usepackage[latin9]{inputenc}
\usepackage{amsmath}
\usepackage{amssymb}
\usepackage{graphicx}

\makeatletter

\usepackage{epsfig}\usepackage{amsfonts}\usepackage{array}\usepackage{dcolumn}\usepackage{psfrag}\usepackage{amsfonts}\usepackage{accents}\usepackage{booktabs}
\usepackage{cite}\usepackage{subfigure}\usepackage[all]{xy}
\usepackage{dsfont}\usepackage{url}\usepackage{xcolor}\usepackage[nolist]{acronym}
\usepackage[footnote,draft,silent,nomargin]{fixme}
\usepackage{algorithm}
\usepackage[noend]{algpseudocode}
\usepackage{flushend}

\makeatother

\begin{document}

\title{A Tractable Model of the LTE Access Reservation Procedure for Machine-Type Communications}

\author{
Jimmy~J.~Nielsen, Dong~Min~Kim, Germ\'an~C.~Madue\~{n}o, Nuno~K.~Pratas, Petar~Popovski\\
APNET section, Department of Electronic Systems, Aalborg University, Denmark\\
\{jjn,dmk,gco,nup,petarp\}@es.aau.dk\\%
}

\maketitle
\begin{abstract} 
A canonical scenario in Machine-Type Communications (MTC) is the one featuring a large
number of devices, each of them with sporadic traffic. Hence, the number of served
devices in a single LTE cell is not determined by the available aggregate rate, but
rather by the limitations of the LTE access reservation protocol. Specifically, the
limited number of contention preambles and the limited amount of uplink grants per
random access response are crucial to consider when dimensioning LTE networks for MTC.
We propose a low-complexity model of LTE's access reservation protocol
that encompasses these two limitations and allows us
to  evaluate the outage probability at click-speed. The model is based chiefly on
closed-form expressions, except for the part with the feedback impact of
retransmissions, which is determined by solving a fixed point equation. Our model
overcomes the incompleteness of the existing models that are focusing solely on the
preamble collisions. A comparison with the simulated LTE access reservation procedure
that follows the 3GPP specifications, confirms that our model provides an accurate
estimation of the system outage event and the number of supported MTC devices.


\end{abstract}


\section{Introduction}

\begin{figure}[t]
    \centering
    \includegraphics[angle=-90,width=0.89\columnwidth]{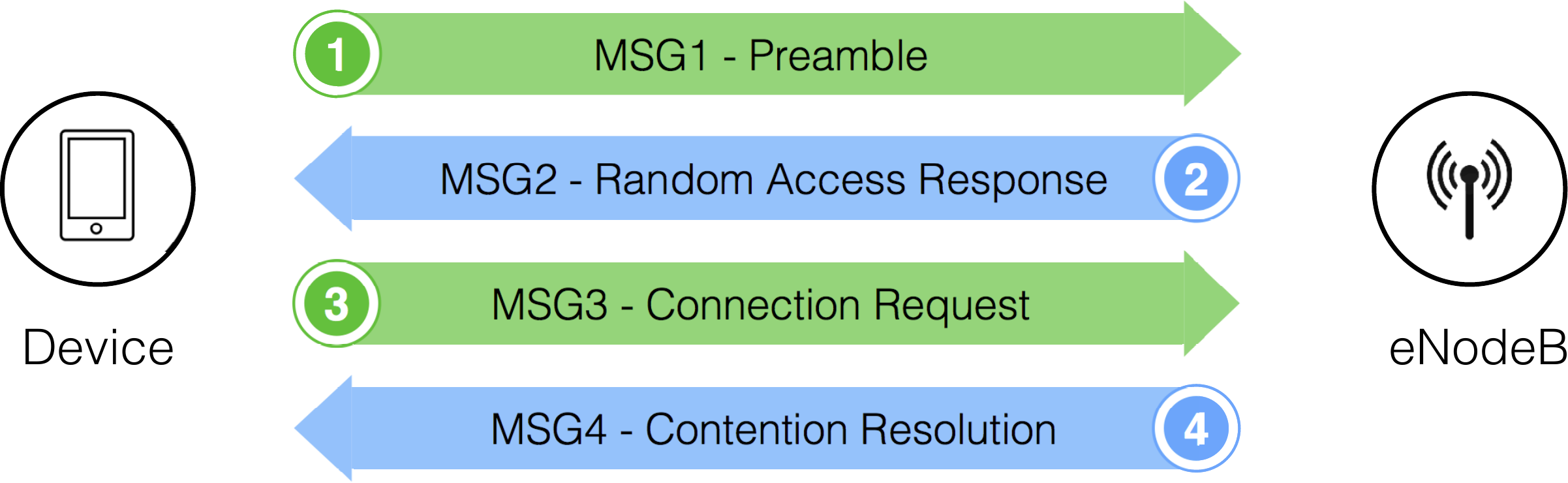}
    \caption{Message exchange between a device and the eNodeB during the LTE random access procedure.}
    \label{fig:LTE_msg1-msg4}
\end{figure}

Machine-Type Communication (MTC) is commonly characterized by a large number of cellular
devices that are active sporadically, where a large number of devices may activate in a
correlated way due to a sensed physical phenomenon (e.g., a power outage in the smart
grid). In more traditional human-centric traffic where the associated payloads are
relatively large, a small number of active devices can cause the network to become in
outage mainly due to the lack of available resources for data transmission. In contrast,
the associated payloads are relatively small in MTC such that the division of the
aggregate available data rate with the small data rate required by each Machine-Type
Device (MTD) leads to the conclusion that the system can support a vast number of MTDs.
Recent studies have shown that such a conclusion is misleading: the network still becomes
in outage, not being able to provide access to the MTDs, despite plenty of available
resources to support a massive number of MTDs. Here the culprit in the limited number of
supported devices, is not the available resources as in human-centric traffic, but
instead the bottlenecks in the access reservation protocol~\cite{Madueno2014}.
Specifically in LTE, the Access Reservation protocol that is outlined in
Fig.~\ref{fig:LTE_msg1-msg4} has two limitations that unveil with MTC. The first is in
\emph{MSG 1}, where only a limited number of preambles can be used to signal a sporadic
request for uplink resources to the eNodeB, in the RACH phase. The second is in \emph{MSG
2}, where a bottleneck may be caused by the limited amount of feedback resources in the
access granting (AG) phase.

In the literature, analytical models of the preamble collision probability have already been considered in standardization documents \cite{3gpp2006R1-061369,3gpp2011R2-112198,3gpp2011tr37868} and scientific papers
\cite{cheng2012rach,ubeda2012lte}.
In \cite{karupongsiri2014random} the preamble collision probability is used to estimate the success probability of transmission attempts.
However, we have found that existing models are incomplete and inaccurate and in this paper we introduce a superior model that closely matches the system outage breaking point of the detailed simulation.

The second limitation in the AG phase has been considered separate from collisions in \cite{yang2014m2m} for bursty arrivals following the Beta
distribution, which is a valuable result for situations where many
alarm messages are sent simultaneously.
In \cite{ubeda2012lte} the authors present an approach to cell planning and adaptation of PRACH resources that only takes
into account the preamble collisions. As we show in this paper, the
AG phase is a limiting factor before the amount of preamble collisions becomes
an issue, since the impact of occasional collisions is effectively diminished with
retransmissions.
In \cite{osti2014analysis} the authors present an analysis accounting for preamble collision and the AG phase, which however does not consider retransmissions.

\begin{figure*}[t]
    \centering
    \includegraphics[width=0.7\linewidth]{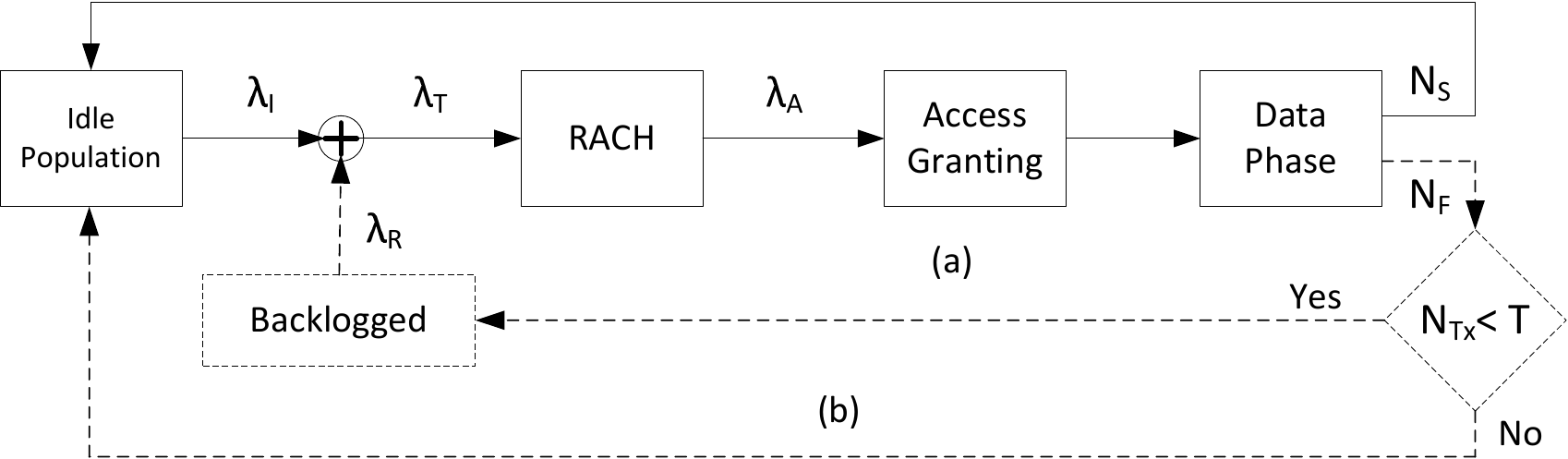}
    \protect\protect\protect\protect
    \caption{Flow diagram of LTE access reservation protocol: (a) one-shot transmission model; (b) $m$-Retransmissions model (dashed lines).}
    \label{fig:LTE_diagram}
\end{figure*}

In this paper we propose an analytical model of the transmission failure probability in
an LTE cell for sporadic uplink transmissions carried over the LTE random access channel.
The proposed model captures the features of the existing access reservation protocol in 
LTE, meaning that we are not proposing a new access protocol rather introducing a tool 
for analysis of the existing LTE access reservation protocol.
The purpose of the proposed model is to be able to estimate the capacity in terms of the
number of terminals or RACH arrival density that can be supported by LTE in a given
configuration while accounting for retransmissions as well as modeling the bottlenecks
that appear in the contention phase and the AG phase. This is a major contribution of the
paper, as the existing models do not capture these bottlenecks. Three other contributions
are: 1) an iterative procedure to determine the impact of retransmissions using a Markov
chain model of the retransmission and backoff procedure; 2) analytical derivations of the
metrics based on a Markov chain, thereby achieving an analytical model that can be
evaluated at click-speed. 3) analysis of the protocol breaking point using ever
increasing access loads to the network.
%
%

Initially we present the system model and assumptions in Section~\ref{sec:system_model},
whereafter in Section~\ref{sec:arp_model} we present the proposed analytical model of the
access reservation bottlenecks in LTE. The proposed model is compared numerically to
simulation results and other models from the literature in Section~\ref{sec:results}, and
finally the conclusions are given in Section~\ref{sec:conclusion}.

\section{System Model}
\label{sec:system_model}
We focus our analysis on a single LTE cell, with $N$ MTDs, also called machine-type User Equipment (UE). We assume that
the MTC applications associated with these MTDs, generate new uplink transmissions with an
aggregate rate $\lambda_\text{I}$, as depicted in Fig.~\ref{fig:LTE_diagram}.
That is, $\lambda_\text{I}=\sum_{i=1}^K \lambda^\text{app}_i$, where $\lambda^\text{app}_i$ is the transmission
generation rate of the $i$th out of $K$ MTC applications running on the UEs. We assume this aggregate rate
follows a Poisson distribution with rate $\lambda_\text{I}$. For each new data transmission,
up to $m$ retransmissions are allowed, resulting in a maximum of $m\!+\!1$ allowed
transmissions. When these transmissions fail and retransmissions occur, then an
additional stress is put on the access reservation protocol, since the rate of retransmissions
$\lambda_\text{R}$ adds to the total rate $\lambda_\text{T}$.

As shown on Fig.~\ref{fig:LTE_diagram}, we split the access reservation model
into two parts: (i) the one-shot transmission part in
Fig.~\ref{fig:LTE_diagram}(a) that models the bottlenecks at each stage of the access
reservation protocol; (ii) the $m$-retransmission part in
Fig.~\ref{fig:LTE_diagram}(b), where a finite number of retransmissions and backoffs are
modeled. We focus our analysis on MTC, for which the traffic is characterized by having a very small payload. Therefore, in the one-shot transmission, depicted in
Fig.~\ref{fig:LTE_diagram}(a), we assume that the RACH and access granting phases are the
system bottlenecks. In other words, we assume that the network has enough data resources
to deliver the serviced MTC traffic.

\subsection{LTE Access Reservation Protocol} 
\label{sub:lte_access_reservation_protocol}

The uplink resources in LTE for frequency division duplexing (FDD) are divided into  time and frequency units denoted resource blocks (RBs)~\cite{3GPPTS36.201}.
The time is divided in frames, where every frame has ten subframes, each subframe of duration $t_s = 1$~ms.
The system bandwidth determines the number of RBs per subframe that ranges between 6~RBs and 100~RBs.
The number of subframes between two consecutive random access opportunities (RAOs) denoted $\delta_\text{RAO}$ varies between 1 and 20.
Every RAO occupies 6~RBs and up to 1 RAO per subframe is allowed.

The LTE random access follows the access reservation principle meaning that devices must
contend for uplink transmission resources in a slotted ALOHA fashion within a
RAO~\cite{3GPPTS36.321,3GPPTS36.213}. As shown in Fig.~\ref{fig:LTE_msg1-msg4}, the
access procedure consists of the exchange of four different messages between a UE and
the eNodeB. The first message (MSG~1) is a randomly selected preamble sent in the first
coming RAO. In Fig.~\ref{fig:LTE_diagram}(a) the intensity of UE requests leading to preamble activations is
represented by $\lambda_\text{T}$. LTE has 64 orthogonal preambles, where only $d=54$ are typically
available for contention among devices, since the rest are reserved for timing alignment. Commonly, the eNodeB can only detect which
preambles have been activated but not if multiple activations (collisions) have occured.
This assumption holds in small cells~\cite[Sec. 17.5.2.3]{sesia2011lte}, and refers to
the worst-case scenario where the detected preamble does not reveal anything about how many users are simultaneously sending that preamble\footnote{When the cell size is more than twice the
distance corresponding to the maximum delay spread, the eNodeB may be able to
differentiate the preamble has been activated by two or more users, but only if the users
are separable in terms of the Power Delay Profile \cite{sesia2011lte,6530825}.}. In other
words, the preamble collision is not detected at MSG~1.

Thereafter, in MSG~2, the eNodeB returns a random access response (RAR) to all detected preambles.
The intensity of activated preambles is in Fig. \ref{fig:LTE_diagram}(a) represented by $\lambda_\text{A}$, where $\lambda_\text{A} \leq \lambda_\text{T}$ since in a preamble collision only 1 preamble is activated.
The contending devices listen to the downlink channel, expecting MSG~2 within $t_{\mathrm{RAR}}$.
It should be noted that typically a maximum of 3 RAR messages per subframe can be sent by the eNodeB \cite{yang2014m2m}.
If no MSG~2 is received and the maximum of $T$ MSG~1 transmissions has not been reached, the device backs off and restarts the random access procedure after a randomly selected backoff time within the interval $t_r\in[0,W_\text{c}] \cap \mathbb{Z}^+$, where $W_\text{c}$ is the maximum backoff time.
If received, MSG~2 includes uplink grant information, that indicates the RB in which the connection request (MSG~3) should be sent.
The connection request specifies the requested service type, e.g., voice call, data transmission, measurement report, etc.
In case of collision the devices receive the same MSG~2, resulting in their MSG~3s colliding in the RB.

In contrast to the collisions of MSG~1, the eNodeB is able to detect collisions of MSG~3. The eNodeB only replies to the MSG~3s that did not experience collision, by sending message MSG~4, with which the required RBs are allocated or the request is denied in case of insufficient resources. The latter is however unlikely in the case of MTC, due to the small payloads. If the MSG~4 is not received within $t_{\mathrm{CRT}}$ since MSG~1 was sent, the random access procedure is restarted. Finally, if a device does not
successfully finish all the steps of the random access procedure within $m\!+\!1$ MSG~1 transmissions, an outage is declared.


\section{Modeling the Access Reservation Protocol}
\label{sec:arp_model} 

We now go to the analysis of the access reservation procedure. First, we
model the \emph{One-shot transmission} and then extend it to the
\emph{$m$-Retransmissions} model. The numerical results cover the
complete model, as depicted in Fig.~\ref{fig:LTE_diagram}.

\subsection{One-Shot Transmission Model} 
\label{sub:one_shot_transmission_model}

We are interested in characterizing how often a transmission from a UE fails. This happens when the transmission is not successful in both the preamble contention and AG phases, i.e., a request from the UE must not experience a preamble collision and the uplink grant must not become stale and dropped.
We model this as two independent events:
\begin{equation}\label{eq:p_s_one-shot}
	p_\text{f}(\lambda_\text{T}) = 1-\Big(1-p_\text{c}(\lambda_\text{T})\Big)\Big(1-p_\text{e}(\lambda_\text{A})\Big),
\end{equation}
where $p_\text{c}(\lambda_\text{T})$ is the collision probability in the preamble contention phase given UE request rate $\lambda_\text{T}$, while $p_\text{e}(\lambda_\text{A})$ is the probability of the uplink grant being dropped from the AG queue given preamble activation rate $\lambda_\text{A}$.

\subsubsection{Preamble Contention Phase} 
\label{sub:preamble_contention_phase}
We start by computing $p_\text{c}(\lambda_\text{T})$.
Let $d$ denote the number of available preambles ($d=54$).
Let the probability of not selecting the same preamble as one other UE be $1-\frac{1}{d}$.
Then the probability of a UE selecting a preamble that has been selected by at least one other UE, given $N_\text{T}$ contending UEs, is:
\begin{equation}
	\mathbb{P} \left(\mbox{Collision}|N_\text{T}\right) = 1 - \left( 1 - \frac{1}{d}\right)^{N_\text{T}-1}.
\end{equation}
Assuming Poisson arrivals with rate $\lambda_\text{T}$, then:
\begin{align}
	p_\text{c}(\lambda_\text{T}) &=\!\sum_{i=1}^{+\infty}\!\left[ 1 - \left(\! 1 - \frac{1}{d}\right)^{i-1} \! \cdot \mathbb{P}(N_\text{T}\!=\!i , \lambda_\text{T} \cdot \delta_\text{RAO})\right] \\ \nonumber
				  & \leq  1 - \left( 1 - \frac{1}{d}\right)^{\lambda_\text{T} \cdot \delta_\text{RAO}-1},
\end{align}
where $\mathbb{P}(N_\text{T} = i , \lambda_\text{T} \cdot \delta_\text{RAO})$ is the probability mass function of the Poisson distribution with arrival rate $\lambda_\text{T} \cdot \delta_\text{RAO}$. The inequality comes from applying Jensen's inequality to the concave function $1 - \left( 1 - 1/d\right)^x$, where $\lambda_\text{T}$ is the total arrival rate (including retransmissions), and $\delta_\text{RAO}$ is the average number of subframes between RAOs.\footnote{E.g., $\delta_\text{RAO}\!=\!1$ if 10 RAOs per frame and $\delta_\text{RAO}\!=\!5$ if 2 RAOs per frame.}
The computed $p_\text{c}(\lambda_\text{T})$ is thus an upper bound on the collision probability.


\subsubsection{Access Granting Phase} 
\label{sub:access_grant_phase}

The mean number of activated preambles in the contention phase per RAO, is given by $\lambda_\text{A}$.
As discussed in Section~\ref{sec:system_model}, we assume that the eNodeB is unable to discern between preambles that have been activated by a single user and multiple users, respectively.
This will lead to a higher $\lambda_\text{A}$, than in the case where the eNodeB is able to detect the preamble collisions.
The main impact of this assumption is that there will be an increased rate of AG requests, even though part of these correspond to collided preambles, which even if accepted will lead to retransmissions.

The $\lambda_\text{A}$ can be well approximated, while assuming that the selection of each preamble by the contending users is independent, by,
\begin{equation}\label{eq:lambda_t_nodetect}
    \lambda_\text{A} = \left[1 - \mathbb{P} \left(X=0 \right)\right] \cdot d,
\end{equation}
where $\mathbb{P}(X=k)$ is the probability of k successes, which can be well approximated with a Poisson distribution with arrival rate $\lambda_\text{T}/d$, i.e.:
\begin{equation}
    \mathbb{P} \left(X=k \right) \approx \frac{(\lambda_\text{T}/d)^k e^{-\lambda_\text{T}/d}}{k!}.
\end{equation}
%

To compute the outage probability due to the limitation in the AG phase, i.e., due to the maximum number of uplink grants per subframe and a maximum waiting time of $t_\text{RAR}$ subframes, we consider that this subsystem can be modeled as a queuing system.
We assume that the loss probability $p_\text{e}(\lambda_\text{A})$ can be seen as the \emph{long-run fraction of costumers that are lost} in a queuing system with impatient costumers \cite{de1985queueing}.

In LTE, pending uplink grants are served with a deterministic time interval (1 subframe) between each serving slot \cite{yang2014m2m}. A straightforward approach would be to use an M/D/1 model structure, as presented in \cite{de1985queueing}, in order to compute the drop probability.
Unfortunately, the expression to compute $p_\text{e}(\lambda_\text{A})$ for the M/D/1 queue does not have a closed-form solution.
However the equivalent expression for the M/M/1 queue in \cite{de1985queueing} has a closed-form solution, see eq. \eqref{eq:p_e}.
We have compared the results of the two model types and found no noticeable difference in the computed outage numbers in practice.
Thus, in the following we use the M/M/1 model to compute $p_\text{e}(\lambda_\text{A})$:
\begin{align}\label{eq:p_e}
	p_\text{e}(\lambda_\text{A}) &= \frac{(1-\rho) \cdot \rho \cdot \Omega}{1-\rho^2 \cdot \Omega}, \text{ with } \Omega = e^{-\mu \cdot (1-\rho) \cdot \tau_\text{q}}.
\end{align}
where $\rho=\frac{\lambda_\text{A}}{\mu}$ is the queue load, $\mu$ is the number of uplink grants per RAR ($\mu=3$), with $\tau_\text{q} = T_\text{d}-\frac{1}{\mu}$ and $T_\text{d}$ is the max waiting time (in terms of requests) in the uplink grant queue, i.e., $T_\text{d} = \mu \cdot t_\text{RAR}$.

The fact that we are using an M/M/1 model instead of an M/D/1 model, may cause a discrepancy between the simulation and model results when the queue becomes congested ($\rho > 1$).
However, we are interested in the switching point ($\rho=1$) from which we then estimate accurately the outage breaking point, as shown in the results in section \ref{sec:results}.



\subsection{$m$-Retransmissions Model} 
\label{sub:t_retransmissions_model}

When UEs are allowed to make retransmissions the probability of an UE
becoming in outage is the probability that none of the allowed $m\!+\!1$ transmissions
attempts are successful.

When retransmissions are allowed ($m>0$), the total arrival rate $\lambda_\text{T}$ must include
the extra arrivals caused by the UEs' retransmissions. The number of retransmissions $\lambda_\text{R}$ is
however a result of the limit $m$ and transmission error probability $p_\text{f}$, which in turn depends on the
number of retransmissions $\lambda_\text{R}$. This chicken and egg problem can be solved iteratively using a
derivative of the Bianchi model \cite{bianchi2000performance} applied to our system
model. Specifically, we are using a model adapted to LTE, with a structure similar to the one presented in
\cite{yang2012performance}. The following derivations of the number of
transmissions and outage probabilities have, to the best of our knowledge, not been
presented previously.
\begin{figure}[t]
	\centering \includegraphics[width=9cm]{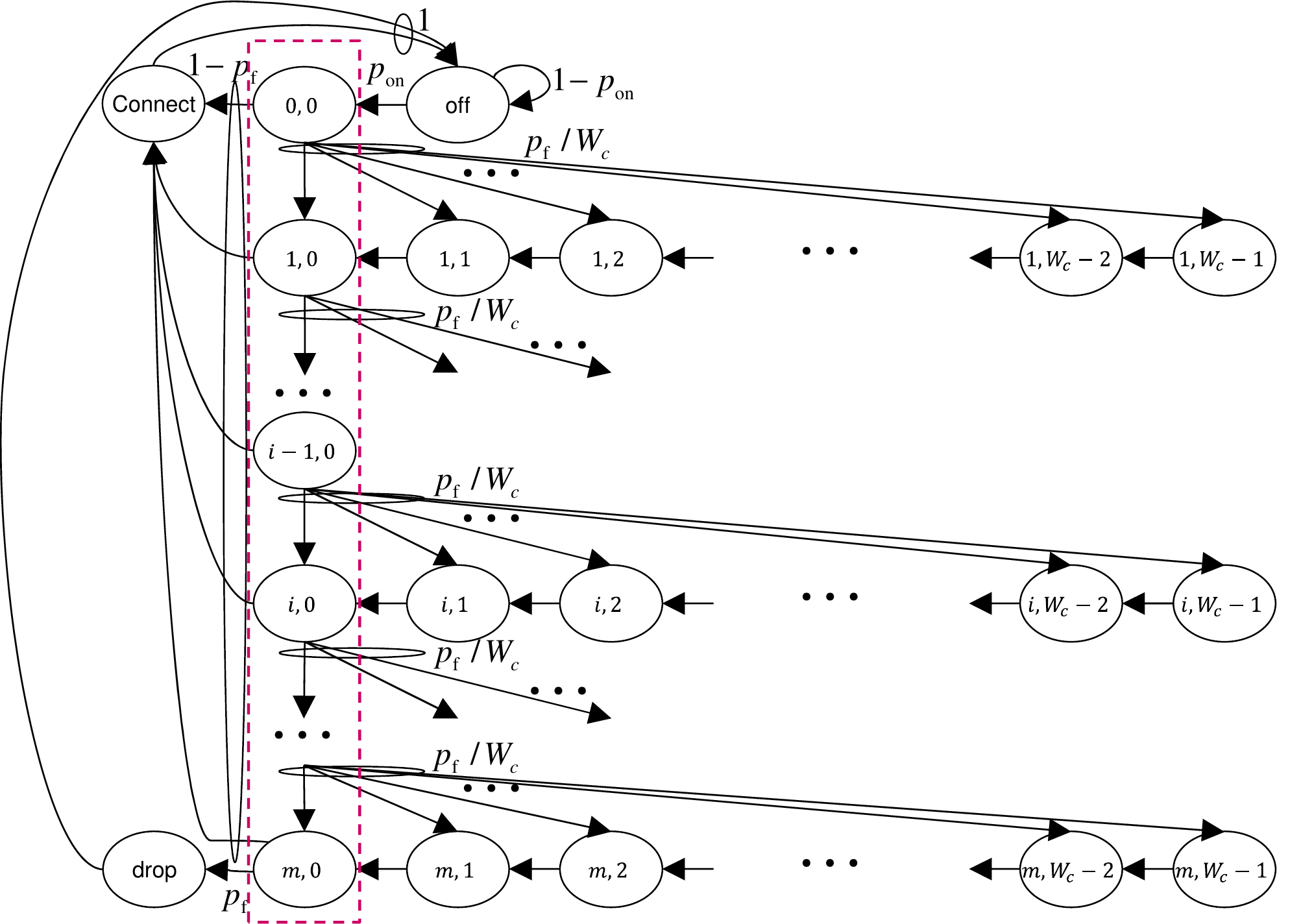} \protect\protect\protect\protect\caption{Markov Chain backoff model to estimate number of required transmissions. The states in the red dashed box are used to calculate $N_\text{TX}$.}
 	\label{fig:mc_backoff_model}
\end{figure}

The mean number of required transmissions $N_\text{TX}$ and outage probability $P_\text{outage}$, are computed with help of the
Markov chain model depicted in Fig.~\ref{fig:mc_backoff_model}. In the Markov chain
model, the state index $\{i, k\}$ denotes the $i$th transmission attempt stage and
$k$th backoff counter. The number of allowed retransmissions is then given by $m$.

Whenever the one-shot transmission is successful, depicted in Fig.~\ref{fig:LTE_diagram}(a), the UE enters the \emph{connect} state:
\begin{equation*}
	\mathbb{P}\left( {\left. {{\text{connect}}} \right|i,0} \right) = {1 - p_\text{f}},\quad0 \leq i \leq m.
\end{equation*}
Where, $p_\text{f}$ is short for $p_\text{f}(\lambda_\text{T})$. Whenever the one-shot access fails, the UE increases the backoff counter:
\begin{equation*}
	\mathbb{P}\left( {\left. {i,k} \right|i - 1,0} \right) = \frac{p_\text{f}}{W_c},
\end{equation*}
where $0 \leq k \leq W_c - 1$ and $1 \leq i \leq m$.

At the last stage of the Markov chain, the UE enters the \emph{drop} state if the transmission fails:
\begin{equation*}
	\mathbb{P}\left( {\left. {{\text{drop}}} \right|m,0} \right) = p_\text{f}(\lambda_\text{T}).
\end{equation*}
The UE enters the \emph{off}
state after the \emph{connect} or the \emph{drop} states, with probability:
\begin{equation*}
	\mathbb{P}\left( {\left. {{\text{off}}} \right|{\text{drop}}} \right) = \mathbb{P}\left( {\left. {{\text{off}}} \right|{\text{connect}}} \right) = 1.
\end{equation*}

From the \emph{off} state, the node enters the first transmission state $\{0,0\}$ with
probability $p_\text{on}$:
\begin{equation*}
	\mathbb{P}\left( {\left. {0,0} \right|{\text{off}}} \right) = {p_\text{on}}.
\end{equation*}
where the probability $p_\text{on}$ is defined as $p_\text{on} =
1-e^{-\lambda_{\text{I}}}$.

Let $b_{i,k}$ be the steady state probability that a UE is at state $\{i,k\}$.
Then $b_{i,k}$ can be derived as:
\begin{align}\label{eq:bik}
  {b_{i,k}} \!&=\! \frac{{W_\text{c} \!-\! k}}{W_\text{c}}{p_\text{f}}{b_{i \!-\! 1,0}} \!=\! \frac{{W_\text{c} \!-\! k}}{W_\text{c}}p_\text{f}^i{b_{0,0}}
   \!=\!\frac{{W_\text{c} \!-\! k}}{W_\text{c}}p_\text{f}^i{p_{{\text{on}}}}{b_{{\text{off}}}},
\end{align}
for $1 \leq i \leq m$ and $0 \leq k \leq W_\text{c}-1$.

Let $b_{\text{connect}}$ be the steady
state probability that a node is at \emph{connect} state:
\begin{align*}
  {b_{{\text{connect}}}} &= \sum\limits_{i = 0}^m {\left( {1 - {p_\text{f}}} \right){b_{i,0}}}  = \sum\limits_{i = 0}^m {\left( {1 - {p_\text{f}}} \right)p_\text{f}^i{p_{{\text{on}}}}{b_{{\text{off}}}}}  \\
   &= \left( {1 - p_\text{f}^{m + 1}} \right){p_{{\text{on}}}}{b_{{\text{off}}}}.
\end{align*}

By imposing the probability normalization condition, as detailed in Appendix
\ref{app:b_off}, we find ${b_{{\text{off}}}}$ as:
\begin{equation*}
	{b_{{\text{off}}}} = \frac{{2\left( {1 - {p_\text{f}}} \right)}}{{2\left( {1 - {p_\text{f}}} \right)\left( {1 + 2{p_{{\text{on}}}}} \right) + {p_{{\text{on}}}}\left( {W_\text{c} + 1} \right){p_\text{f}}\left( {1 - p_\text{f}^m} \right)}}.
\end{equation*}

Since a transmission will eventually either finish successfully in the \emph{connect}
state or unsuccessfully in the \emph{drop} state, the outage probability can be computed
as:
\begin{align}
	P_\text{outage} = \frac{b_\text{drop}}{b_\text{drop}+b_\text{connect}} = p_\text{f}^{m + 1},
\end{align}
where $b_\text{drop}$ and $b_\text{connect}$, whose derivations are shown in the Appendix~\ref{app:b_off}, can be computed as:
\begin{align}
   {b_{{\text{connect}}}} \!=\! \frac{{2\left( {1 \!-\! {p_\text{f}}} \right)\left( {1 \!-\! p_\text{f}^{m + 1}} \right){p_{{\text{on}}}}}}{{2\left( {1 \!-\! {p_\text{f}}} \right)\left( {1 \!+\! 2{p_{{\text{on}}}}} \right) \!+\! {p_{{\text{on}}}}\left( {W_\text{c} \!+\! 1} \right){p_\text{f}}\left( {1 \!-\! p_\text{f}^m} \right)}}, \\
  {b_{{\text{drop}}}} = \frac{{2\left( {1 \!-\! {p_\text{f}}} \right)p_\text{f}^{m + 1}{p_{{\text{on}}}}}}{{2\left( {1 \!-\! {p_\text{f}}} \right)\left( {1 \!+\! 2{p_{{\text{on}}}}} \right) \!+\! {p_{{\text{on}}}}\left( {W_\text{c} \!+\! 1} \right){p_\text{f}}\left( {1 \!-\! p_\text{f}^m} \right)}}.
\end{align}
%
%
%

The number of required transmissions can be estimated from the steady state probabilities,
keeping in mind that $b_{i,0}/b_{0,0}$ represents the probability of using $i+1$ or more
transmission attempts to deliver a packet, and $b_{m,0}/b_{0,0}$ is the probability of
using exactly $m\!+\!1$ transmission attempts:
\begin{align}\label{eq:entx}
	N_\text{TX}(\lambda_\text{T}) \!&=\! \frac{\left(\sum\limits_{i=0}^{m-1} (i+1) \cdot (b_{i,0}-b_{i+1,0})\right) + (m+1)\cdot b_{m,0}}{b_{0,0}} \nonumber \\
    \!&=\! \left( {1 \!-\! {p_\text{f}}} \right)\!\!\sum\limits_{i = 0}^{m - 1} \!\!{\left( {i \!+\! 1} \right)\!p_\text{f}^i}  \!+\! \left( {m \!+\! 1} \right)p_\text{f}^m \!=\! \frac{{1 - p_\text{f}^{m + 1}}}{{1 - {p_\text{f}}}}.
\end{align}

From the number of transmissions, the value of $\lambda_\text{T}$ can be solved iteratively using the fixed point equation:
\begin{align}
\lambda_\text{T} = N_\text{TX}(\lambda_\text{T}) \cdot \lambda_\text{I}
= \lambda_\text{I} \frac{1-p_\text{f}(\lambda_\text{T})^{m+1}}{1-p_\text{f}(\lambda_\text{T})}.
\end{align}

\begin{figure*}
    \centering
        \subfigure[Outage probability, $m=0$]{
        \includegraphics[width=0.3\textwidth]{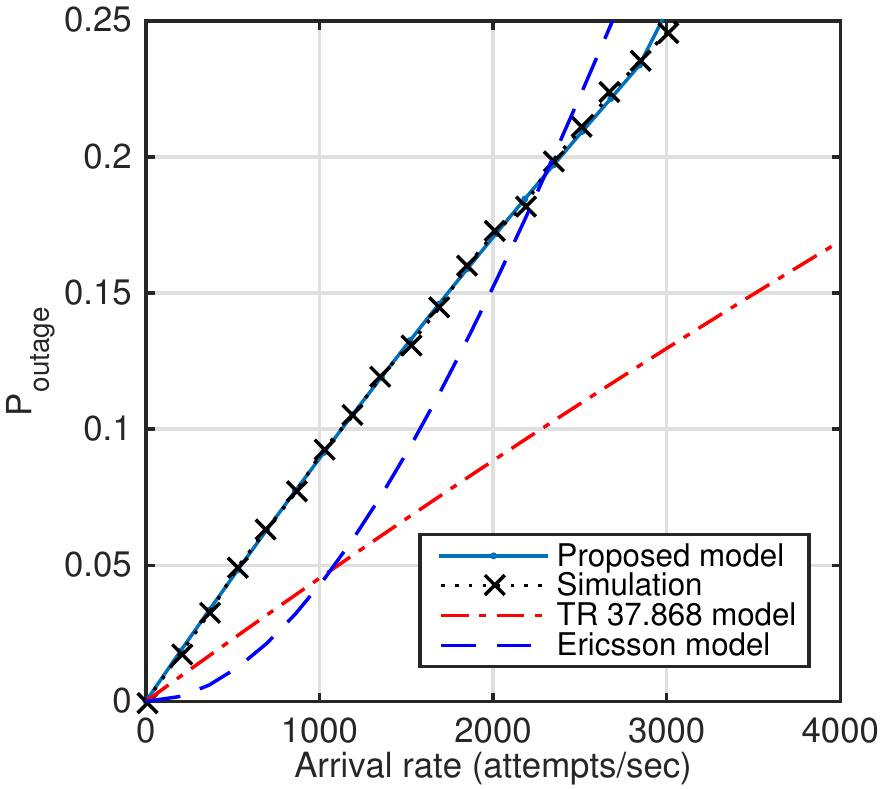}
        \label{fig:outage_RACH_2_ntx_1}
        }
        \subfigure[Expected number of transmissions, $m=9$]{
        \includegraphics[width=0.31\textwidth]{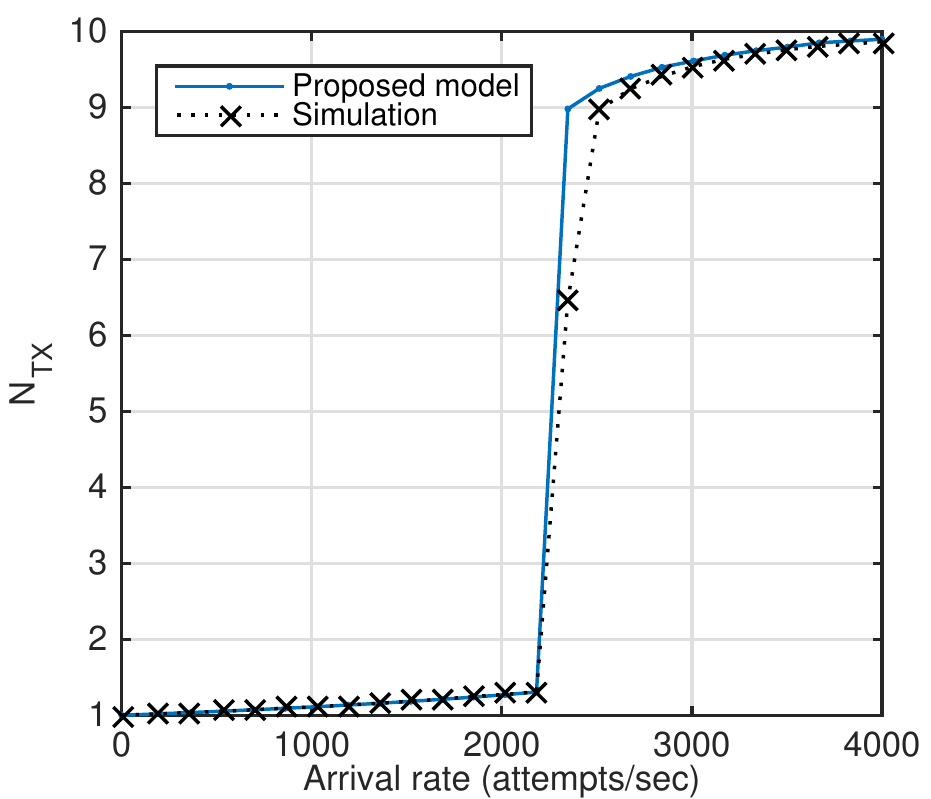}
        \label{fig:entx_RACH_2_ntx_10}
        }
        \subfigure[Outage probability, $m=9$]{
        \includegraphics[width=0.3\textwidth]{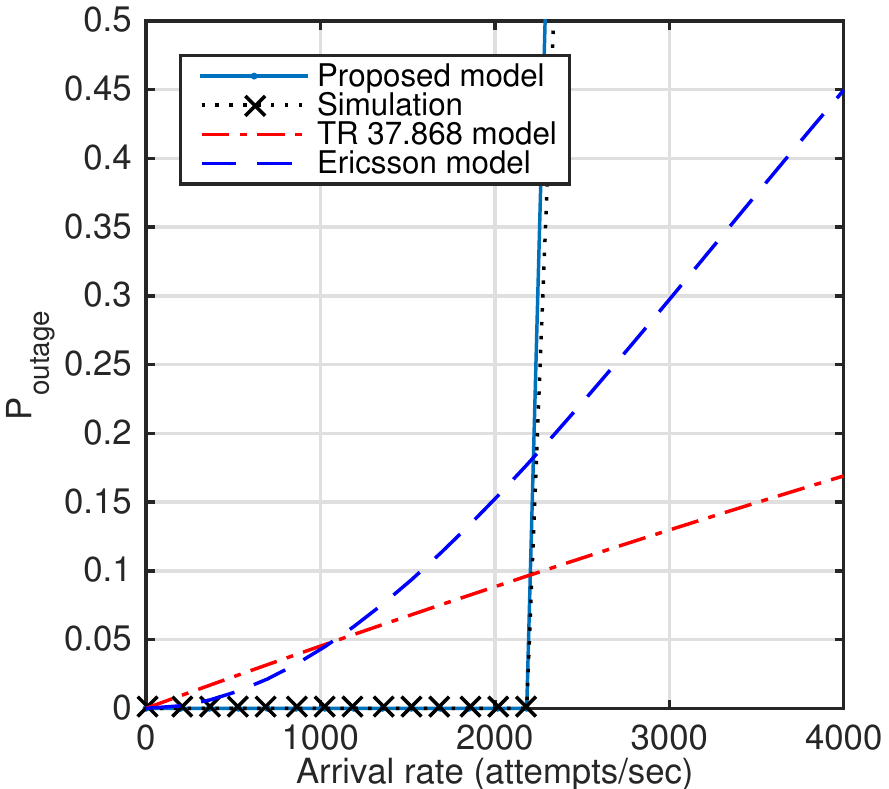}
        \label{fig:outage_RACH_2_ntx_10}
    }
    \caption{Plots for RACH configuration with 2 RAOs per frame ($\delta_\text{RAO}=5$). Ericsson model refers to \cite{ubeda2012lte}.}
    \label{fig:plots_rach_2}
\end{figure*}

\begin{figure*}
    \centering
        \subfigure[Outage probability, $m=0$]{
        \includegraphics[width=0.3\textwidth]{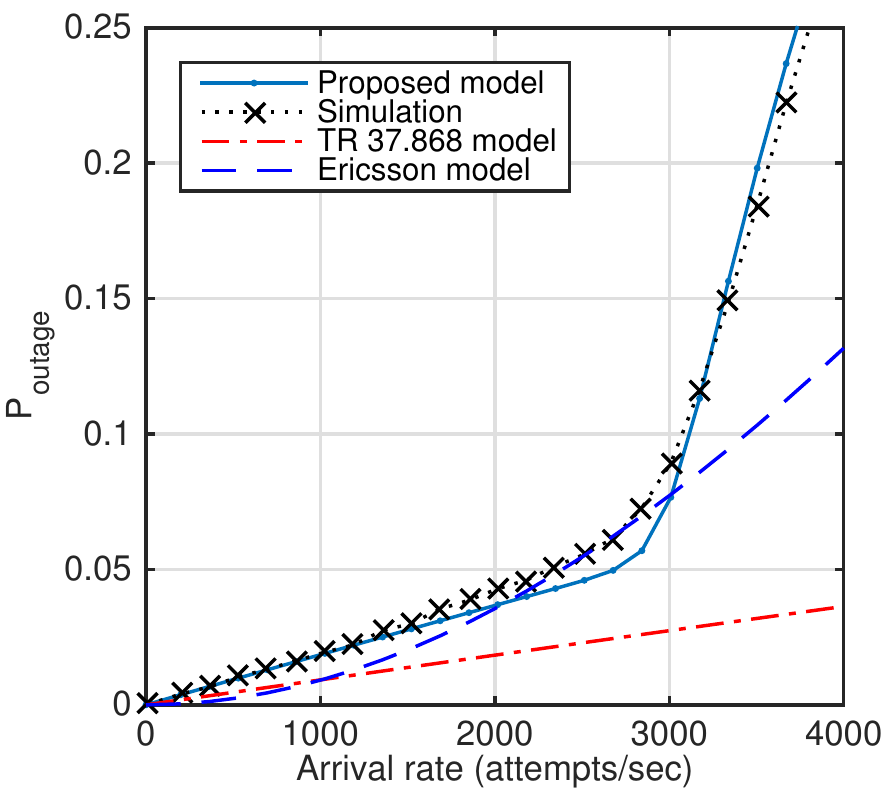}
        \label{fig:outage_RACH_10_ntx_1}
        }
        \subfigure[Expected number of transmissions, $m=9$]{
        \includegraphics[width=0.3\textwidth]{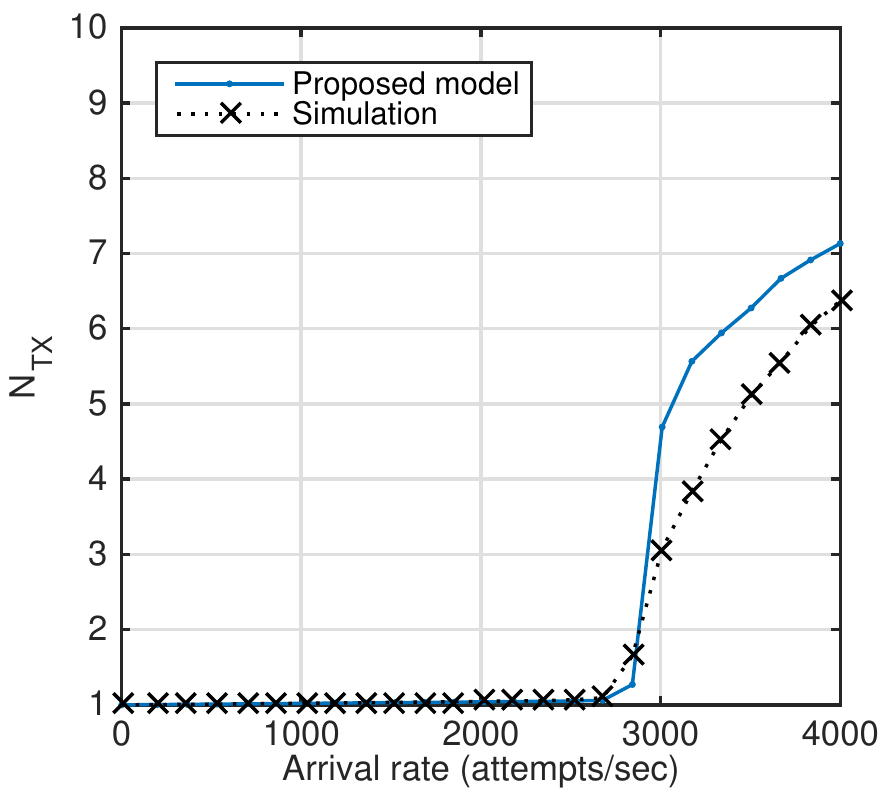}
        \label{fig:entx_RACH_10_ntx_10}
        }
        \subfigure[Outage probability, $m=9$]{
        \includegraphics[width=0.3\textwidth]{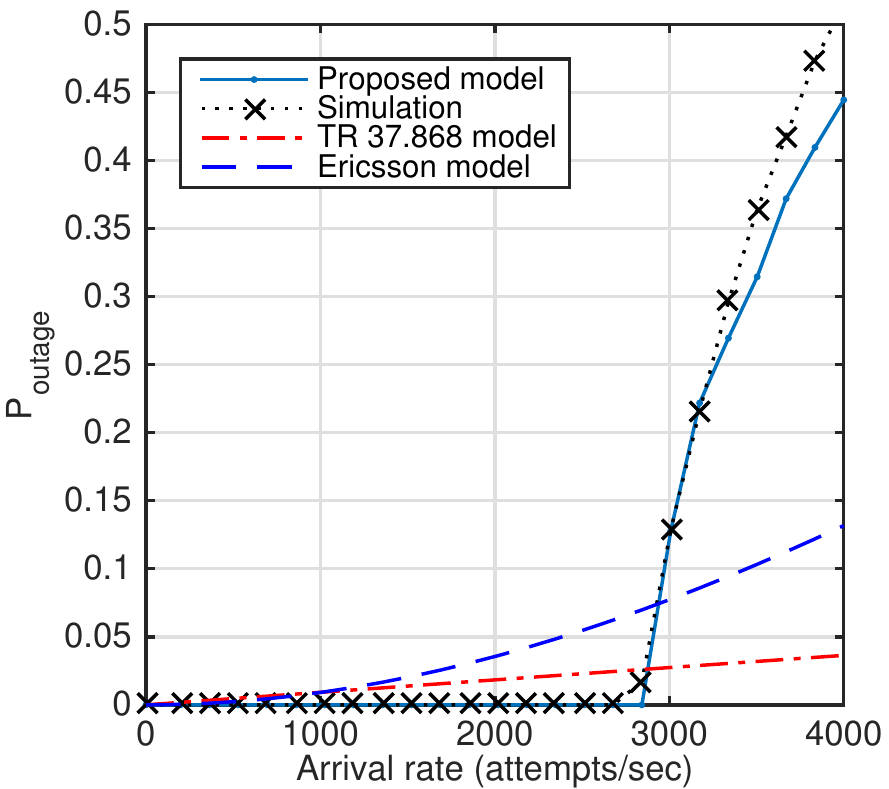}
        \label{fig:outage_RACH_10_ntx_10}
    }
    \caption{Plots for RACH configuration with 10 RAOs per frame ($\delta_\text{RAO}=1$). Ericsson model refers to \cite{ubeda2012lte}.}
    \label{fig:plots_rach_10}
\end{figure*}

For the results presented in Sec. \ref{sec:results} we found that less than 20 iterations were needed to reach convergence (less than 1\% change between consecutive iterations).


\section{Numerical results}
\label{sec:results} 

In our evaluation, we consider two PRACH configurations, namely the typical configuration  with 5 subframes between every RAO \cite{typical} and the configuration with one RAO every subframe.
Further, we consider first the case where only a single transmission is allowed (one-shot, $m\!=\!0$) and then the more realistic configuration of $m\!=\!9$ allowed retransmissions.
The model results are compared with a simulator that implements the full LTE access reservation protocol as defined in~\cite{3GPPTS36.321} and \cite{3GPPTS36.213} given parameters in Table \ref{tab:lte_system_parameters}.
\begin{table}[hb]
	\centering
	\begin{tabular}{l|c}
		\textbf{Parameter} & \textbf{Value} \\ \hline
			Preambles per RAO (d) & 54 \\
			Subframes between RAOs ($\delta_\text{RAO}$) & 1 or 5 \\
			Max number of retransmissions ($m$) & 0 or 9 \\
			Uplink grants per RAR ($\mu$) & 3 \\
			System bandwidth & 5~MHz \\
			eNodeB processing time & 3~ms \\
			MSG~2 window ($t_\text{RAR}$) & 5~ms or 10~ms \\
			Contention time-out ($t_{\mathrm{CRT}}$) & 48~ms \\
			Backoff limit ($W_\text{c}$) & 20~ms \\
			UE processing time & 3~ms
	\end{tabular}
	\caption{LTE simulation and model parameters}
	\label{tab:lte_system_parameters}
    \vspace{-0.6cm}
\end{table}

\subsection{One-shot Transmission ($m\!=\!0$)} 
\label{sub:one_shot_transmission}
In Fig.~\ref{fig:outage_RACH_2_ntx_1} and~\ref{fig:outage_RACH_10_ntx_1} the outage probabilities are depicted for $m=0$.
There, the proposed model has a much better fit to the simulation results than the 3GPP TR 37.868 model~\cite[Sec. B.1]{3gpp2011tr37868} and the Ericsson model in~\cite[eq. (6)]{ubeda2012lte}.
Specifically, in Fig.~\ref{fig:outage_RACH_2_ntx_1} where the preamble collisions are the main error cause, the TR 37.868 and Ericsson models are
much worse than the proposed model.
From Fig.~\ref{fig:outage_RACH_10_ntx_1} it is clear that those models are not accounting for the limited number of uplink grants per RAR that starts to have an impact around $\lambda_\text{I}=2700$ attempts/sec, causing an upward bend in the outage curve.


\subsection{$m\!=\!9$ Retransmissions} 
\label{sub:t_10_retransmissions}

In the typical configuration where retransmissions are allowed, a necessary feature of our model is that it is able to account for the feedback impact of retransmissions on the arrival rate $\lambda_\text{T}$.
An intermediate metric that allows to study this is the number of transmissions per new data packet $N_\text{TX}$.
This is shown in Figs.~\ref{fig:entx_RACH_2_ntx_10} and~\ref{fig:entx_RACH_10_ntx_10}.
In Fig.~\ref{fig:entx_RACH_2_ntx_10} the number of transmissions is
estimated accurately leading to a well-fitting estimation of the outage in
\ref{fig:outage_RACH_2_ntx_10}. For the case of 10 RAOs per frame, the Markov chain model
slightly overestimates the number of transmissions. However, the breaking points in the
curves are the same, meaning that the supported arrival rate in the simulation on
Fig.~\ref{fig:outage_RACH_10_ntx_10} is closely matched by the one in the model.

Finally, the results show that the proposed model is superior to the existing models from the literature, as they do not capture the feedback impact of the retransmissions and are therefore not able to estimate the system outage capacity.

The presented results also reveal an interesting insight in dimensioning the LTE access reservation parameters. Given that there is a $5$ times difference in resource usage for RAOs (2 vs 10 RAOs per frame), the gain in supported arrival rate $\lambda_\text{I}$ is quite modest, increasing from around $\lambda_\text{I}=2250$ to around $\lambda_\text{I}=2800$, i.e., a 25\% increase. In order to further increase the capacity of the system, it is necessary to simultaneously increase the number of RARs per subframe.


\section{Conclusions and Outlook}
\label{sec:conclusion} 
In this paper we have presented a low-complexity, yet accurate model to estimate the
outage capacity of the LTE access reservation protocol for machine-type communications,
where the small payload sizes mean that system resources are typically not the
limiting factor. The model accounts for both contention preamble collisions and the
limited number of uplink grants in the random access response message, as well as the
feedback impact that the resulting retransmissions has on the random access load. For the
considered typical LTE configurations, the model is able to very accurately estimate the
system outage capacity. This puts it forward as a useful tool in system dimensioning, as
it allows to replace time-consuming simulations with click-speed calculations.

Future work should look into how diverse channel conditions and diverse traffic patterns
of users can be efficiently included in the model. While the outage metric is very
important from a planning perspective, other metrics such as access delay or transmission
time would be very relevant to be able to estimate accurately when considering real-time
machine-type communications.


\section*{Acknowledgment}
This work is partially funded by EU, under Grant agreement no. 619437. The SUNSEED
project is a joint undertaking of 9 partner institutions and their contributions are
fully acknowledged. This work has been partially supported by the Danish High Technology
Foundation via the Virtuoso project.

\appendix
The following steps are taken to derive $b_\text{off}$ by imposing the probability normalization condition:
\label{app:b_off}
\begin{align*}
  1 &= {b_{{\text{off}}}} + {b_{{\text{conn}}}} + {b_{{\text{drop}}}} + {b_{0,0}} + \sum\limits_{i = 1}^m {\sum\limits_{k = 0}^{W_\text{c} - 1} {{b_{i,k}}} }  \\
   &= {b_{{\text{off}}}} + 2{p_{{\text{on}}}}{b_{{\text{off}}}} + \sum\limits_{i = 1}^m {\sum\limits_{k = 0}^{W_\text{c} - 1} {\frac{{W_\text{c} - k}}{W_\text{c}}p_\text{f}^i{p_{{\text{on}}}}{b_{{\text{off}}}}} }  \\
   &= {b_{{\text{off}}}} + 2{p_{{\text{on}}}}{b_{{\text{off}}}} + {p_{{\text{on}}}}{b_{{\text{off}}}}\left( {\frac{{W_\text{c} + 1}}{2}} \right){p_\text{f}}\frac{{1 - p_\text{f}^m}}{{1 - {p_\text{f}}}}.
\end{align*}

The derivation of $b_\text{connect}$ is as follows:
\begin{align}
   {b_{{\text{connect}}}} &= 1 \!-\!\, {b_{{\text{off}}}} \!\,-\!\, {p_{{\text{on}}}}{b_{{\text{off}}}}(p_\text{f}^{m + 1} \!-\! 1 \!-\! \left( {\frac{{W_\text{c} \!+\! 1}}{2}} \right){p_\text{f}}\frac{{1 \!-\! p_\text{f}^m}}{{1 \!-\! {p_\text{f}}}}) \nonumber \\
   &=\! \frac{{2\!\left( {1 \!-\! {p_\text{f}}} \right)\!\left( {1 \!-\! p_\text{f}^{m + 1}} \right){p_{{\text{on}}}}}}{{2\!\left( {1 \!-\! {p_\text{f}}} \right)\!\left( {1 \!+\! 2{p_{{\text{on}}}}} \right) \!+\! {p_{{\text{on}}}}\!\left( {W_\text{c} \!+\! 1} \right){p_\text{f}}\left( {1 \!-\! p_\text{f}^m} \right)}}.
 \end{align}

The derivation of $b_\text{drop}$ is as follows:
\begin{align}
{b_{{\text{off}}}} &= \left(1-p_{\text{on}}\right){b_{{\text{off}}}} + {b_{{\text{drop}}}} + {b_{{\text{connect}}}} \nonumber \\
{b_{{\text{drop}}}} &=  p_{\text{on}}{b_{{\text{off}}}} - {b_{{\text{connect}}}} \nonumber \\
   &=\! \frac{{2\left( {1 \!-\! {p_\text{f}}} \right)p_\text{f}^{m + 1}{p_{{\text{on}}}}}}{{2\!\left( {1 \!-\! {p_\text{f}}} \right)\!\left( {1 \!+\! 2{p_{{\text{on}}}}} \right) \!+\! {p_{{\text{on}}}}\!\left( {W_\text{c} \!+\! 1} \right){p_\text{f}}\left( {1 \!-\! p_\text{f}^m} \right)}}.
\end{align}


\begin{thebibliography}{10}
\providecommand{\url}[1]{#1}
\csname url@samestyle\endcsname
\providecommand{\newblock}{\relax}
\providecommand{\bibinfo}[2]{#2}
\providecommand{\BIBentrySTDinterwordspacing}{\spaceskip=0pt\relax}
\providecommand{\BIBentryALTinterwordstretchfactor}{4}
\providecommand{\BIBentryALTinterwordspacing}{\spaceskip=\fontdimen2\font plus
\BIBentryALTinterwordstretchfactor\fontdimen3\font minus
  \fontdimen4\font\relax}
\providecommand{\BIBforeignlanguage}[2]{{%
\expandafter\ifx\csname l@#1\endcsname\relax
\typeout{** WARNING: IEEEtran.bst: No hyphenation pattern has been}%
\typeout{** loaded for the language `#1'. Using the pattern for}%
\typeout{** the default language instead.}%
\else
\language=\csname l@#1\endcsname
\fi
#2}}
\providecommand{\BIBdecl}{\relax}
\BIBdecl

\bibitem{Madueno2014}
G.~{Corrales Madue\~no}, C.~Stefanovic, and P.~Popovski, ``{Reengineering
  {GSM/GPRS} Towards a Dedicated Network for Massive Smart Metering},'' in
  \emph{Proc. of the IEEE Internation Conference on Smart Grid Communications
  (SmartGridComm 2014)}, Nov. 2014.

\bibitem{3gpp2006R1-061369}
{3GPP}, ``R1-061369: {LTE} random-access capacity and collision probability,''
  Tech. Rep. RAN1\#45, May 2006.

\bibitem{3gpp2011R2-112198}
------, ``{R2-112198: Clarification on the discussion of {RACH} Collision
  Probability},'' Tech. Rep. RAN2\#73bis, April 2011.

\bibitem{3gpp2011tr37868}
------, ``{TR} 37.868: Study on {RAN} improvements for machine-type
  communications, rel. 11,'' Tech. Rep., September 2011.

\bibitem{cheng2012rach}
R.-G. Cheng, C.-H. Wei, S.-L. Tsao, and F.-C. Ren, ``{{RACH} Collision
  Probability for Machine-Type Communications},'' in \emph{Proc. of the IEEE
  Vehicular Technology Conference (VTC Spring 2012)}, May 2012.

\bibitem{ubeda2012lte}
C.~Ubeda, S.~Pedraza, M.~Regueira, and J.~Romero, ``{{LTE} {FDD} physical
  random access channel dimensioning and planning},'' in \emph{Proc. of the
  IEEE Vehicular Technology Conference (VTC Fall 2012)}, Sept. 2012.

\bibitem{karupongsiri2014random}
C.~Karupongsiri, K.~S. Munasinghe, and A.~Jamalipour, ``Random access issues
  for smart grid communication in {LTE} networks,'' in \emph{Proc. of the
  International Conference on Signal Processing and Communication Systems
  (ICSPCS 2014)}, Dec. 2014.

\bibitem{yang2014m2m}
B.~Yang, G.~Zhu, W.~Wu, and Y.~Gao, ``{{M2M} Access Performance in {LTE-A}
  System},'' \emph{Transactions on Emerging Telecommunications Technologies},
  vol.~25, no.~1, pp. 3--10, Jan. 2014.

\bibitem{osti2014analysis}
P.~Osti, P.~Lassila, S.~Aalto, A.~Larmo, and T.~Tirronen, ``{Analysis of
  {PDCCH} performance for {M2M} traffic in {LTE}},'' \emph{IEEE Trans. Veh.
  Technol.}, vol.~63, no.~9, pp. 4357--4371, Nov. 2014.

\bibitem{3GPPTS36.201}
{3GPP}, ``{TS} 36.201 {E-UTRA} {LTE} physical layer; general description,''
  Tech. Rep., 2015.

\bibitem{3GPPTS36.321}
------, ``{TS} 36.321 {E-UTRA} medium access control ({MAC}) protocol
  specification,'' Tech. Rep., 2015.

\bibitem{3GPPTS36.213}
------, ``{TS} 36.213 {E-UTRA} physical layer procedures,'' Tech. Rep., 2015.

\bibitem{sesia2011lte}
S.~Sesia, I.~Toufik, and M.~Baker, \emph{{LTE-The UMTS Long Term Evolution:
  From Theory to Practice}}.\hskip 1em plus 0.5em minus 0.4em\relax Wiley,
  2011.

\bibitem{6530825}
H.~Thomsen, N.~Pratas, C.~Stefanovic, and P.~Popovski, ``{Analysis of the {LTE}
  Access Reservation Protocol for Real-Time Traffic},'' \emph{IEEE Commun.
  Lett.}, vol.~17, no.~8, pp. 1616--1619, Aug. 2013.

\bibitem{de1985queueing}
A.~G. De~Kok and H.~Tijms, ``A queueing system with impatient customers,''
  \emph{Journal of Applied Probability}, vol.~22, no.~3, pp. 688--696, Sept.
  1985.

\bibitem{bianchi2000performance}
G.~Bianchi, ``{Performance Analysis of the {IEEE} 802.11 Distributed
  Coordination Function},'' \emph{IEEE J. Select. Areas Commun.}, vol.~18,
  no.~3, pp. 535--547, Mar. 2000.

\bibitem{yang2012performance}
X.~Yang, A.~Fapojuwo, and E.~Egbogah, ``{Performance Analysis and Parameter
  Optimization of Random Access Backoff Algorithm in {LTE}},'' in \emph{Proc.
  of the IEEE Vehicular Technology Conference (VTC Fall 2012)}, Sept. 2012.

\bibitem{typical}
3GPP, ``{MTC simulation assumptions for RACH performance evaluation},'' {3rd
  Generation Partnership Project (3GPP)}, TR {R2-105212}, Aug. 2010.
\end{thebibliography}
\end{document}